\documentclass[amsmath,amssymb,superscriptaddress,aip,apl,lengthcheck] {revtex4-1}%

\usepackage{graphics}
\usepackage{bm}
\usepackage[latin1]{inputenc}
\usepackage{textcomp}

\begin{document}

\newcommand{\SiO}{SiO$ _2$}
\newcommand{\LM}{optical microscope}

\title{Imaging ellipsometry of graphene}

\author{Ulrich Wurstbauer}
	\email{Ulrich.Wurstbauer@physik.uni-regensburg.de }
	\altaffiliation[\newline present address:]{ Department of Physics, Columbia University, USA}
	\affiliation{Institut für Experimentelle und Angewandte Physik, Universität Regensburg, 93040 Regensburg, Germany}

\author{Christian Röling}
	 \affiliation{Accurion GmbH, Stresemannstr.30, 37079 Göttingen, Germany}

\author{Ursula Wurstbauer}
	\altaffiliation[present address:]{ Department of Physics, Columbia University, USA}
	\affiliation{Institut für Experimentelle und Angewandte Physik, Universität Regensburg, 93040 Regensburg, Germany}

\author{Werner Wegscheider}
	\altaffiliation[present address:]{ Solid State Physics Laboratory, ETH Zurich, 8093 Zurich, Switzerland}
	\affiliation{Institut für Experimentelle und Angewandte Physik, Universität Regensburg, 93040 Regensburg, Germany}

\author{Matthias Vaupel}
	\affiliation{Accurion GmbH, Stresemannstr.30, 37079 Göttingen, Germany}

\author{Peter H. Thiesen}
	\affiliation{Accurion GmbH, Stresemannstr.30, 37079 Göttingen, Germany}

\author{Dieter Weiss}
	\affiliation{Institut für Experimentelle und Angewandte Physik, Universität Regensburg, 93040 Regensburg, Germany}

\date{\today}

\begin{abstract}
Imaging ellipsometry studies of graphene on \SiO/Si and crystalline GaAs are presented. We demonstrate that imaging ellipsometry is a powerful tool to detect and characterize graphene on any flat substrate. Variable angle spectroscopic ellipsometry is used to explore the dispersion of the optical constants of graphene in the visible range with high lateral resolution.  In this way the influence of the substrate on graphene's optical properties can be investigated.
\end{abstract}


\maketitle\noindent
Graphene is a two-dimensional crystalline solid consisting of only one atomic layer of hexagonally arranged carbon atoms \cite{Neto-Geim-RevModPhys}. To the fascinating properties of this two dimensional lattice belong a record electron/hole mobility at room temperature and charge carriers behaving at low energies like massless Dirac fermions resulting in distinct transport properties such as half integer quantum Hall effect \cite{Novoselov_2d-massless-DiracFermions,Zhang-Kim_nature} and Klein tunneling \cite{Young-KleinTunnelling}. Graphene has also a high potential for devices in various fields including e.g. mechanically very 'robust' transparent electrodes for touch screens, solar cells, photo detectors, nano-electronics and high frequency devices \cite{30Inch, Ferrari-OpticsReview,IBM-HighFreqTransistor}. For the latter a combination with GaAs, currently used for high frequency applications, seems to be promising. However, the detection and characterization of graphene on GaAs-based materials has been reported to be very time-consuming\cite{substratAPL} or limited to special layered GaAs/AlGaAs heterostructures\cite{AhlersAPL-SiO2_like_GaAs}. It is well known that substrate and environment significantly influence graphene's electrical properties \cite{CoryDean-BN-2010,Bolotin-1st-suspended-SolidStateCom,PRL-PbZrTi}. Further, the influence of the substrate and environment on the optical properties of graphene has theoretically been predicted \cite{EllipsometryGrapheneTheory_MeeraSetlur}. As recently reported by Kravets et al.\cite{GeimEllipsometry-PRB-2010} optical constants and the optical dispersion can be extracted from ellipsometric spectra. However, those experiments are limited to extremely large graphitic flakes due to the extended spot size of the light. Picometry allows investigations of the optical properties with a higher lateral resolution but is limited to certain wavelengths\cite{Wang_Picometry}. Graphene oxide layers have already been characterised by imaging ellipsometry\cite{Vaupel-JPhysChemC}.\par\noindent
\begin{figure}
\includegraphics{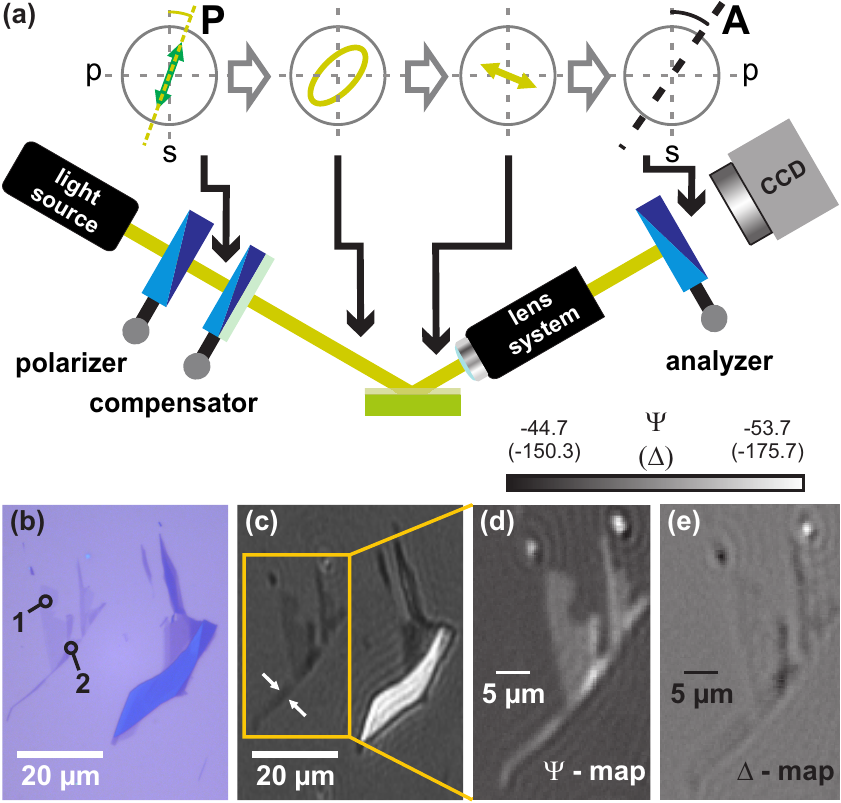}
\caption{(a) Schematic imaging ellipsometry setup. The lens system mounted between sample and analyzer allows imaging with sub-\textmu m lateral resolution. (b) Optical image and (c) imaging ellipsometric intensity image of a sample on \SiO/Si showing regions with graphene monolayer covering up to thin graphite. Numbers in (b) correspond to the layer numbers. Ellipsometric $\Psi$ map (d) and the corresponding $\Delta$ map (e) of the yellow boxed region display graphene mono- and bilayer areas with higher resolution.}
\label{fig:PrinciplesMaps}
\end{figure}
\noindent
In this letter we demonstrate that imaging ellipsometric intensity (IEI) maps, imaging ellipsometry (IE) and imaging variable angle spectroscopic ellipsometry (IVASE) at visible light frequencies are powerful tools to detect and classify graphene flakes and to study their optical properties on a large variety of flat substrates. Due to the high lateral resolution of less than 1~\textmu m of our setup, the optical properties can be mapped over a graphitic flake. Here we report on measurements of exfoliated graphene deposited either on 300~nm amorphous \SiO~on Si or on crystalline GaAs-based substrates, grown by molecular beam epitaxy.\\
The graphene mono- and multilayer samples have been prepared by micromechanical exfoliation of natural graphite as introduced in references\cite{Novoselov_PNAS-2D-Crystals,Novoselov-1stScience,substratAPL}. The flakes were investigated by imaging ellipsometry under ambient conditions at room temperature with a nulling ellipsometer nanofilm\underbar~ep3se from Accurion GmbH \cite{nanofilm} using three different modes as described below. The optical properties are measured for wavelengths ranging from $\lambda = 360$~nm to $\lambda = 1000$~nm (bandwidth $\pm 6$~nm to $\pm 20$~nm). The reflected light from the surface is focused with 20\texttimes~or 50\texttimes~objectives. The latter leads to a 68~\texttimes~79~\textmu m$^2$ field of view and enables a lateral resolution better than 1~\textmu m as demonstrated in Figure \ref{fig:PrinciplesMaps} (c). There, the two arrows mark a width of less than 800~nm. In addition, size, shape, number of layers, morphology and height of the graphene flakes have been determined by a combination of optical, scanning electron (SEM) and atomic force (AFM) microscopy.\par\noindent
In Figure \ref{fig:PrinciplesMaps} (a) the used imaging ellipsometry setup is schematically depicted. The angle of incidence (AOI) is the angle between incident (reflected) light and the sample normal and can be varied. A polarizer polarizes the incoming light linearly which later gets elliptically prepared by a compensator in such a way that the reflected light is again linearly polarized. After passing an analyzer the reflected light is collected by a CCD camera. The lens system enables the high lateral solution. In the upper row of Figure \ref{fig:PrinciplesMaps} (a), the corresponding states of polarizations are sketched. In an appropriately chosen coordinate system the ratio $\rho$ of the perpendicular $p$ and the orthogonal $s$ components of the reflection matrix can be described by $\rho=\frac{E_{out,p}/E_{in,p}}{E_{out,s}/E_{in, s}}=\tan(\psi)\cdot e^{i\Delta}$ with the ellipsometric angles $\Psi$ and $\Delta$ \cite{HandbookOfOpticsII, HandbookOfEllipsometry, Vaupel-JPhysChemC}. $E_{in(out), p(s)}$ denotes the electric field of the incoming (outgoing) light parallel (orthogonal) to the plane of incident as sketched in the top row of Figure \ref{fig:PrinciplesMaps} (a).  In the IEI mode, the angle between polarizer and analyzer as well as the AOI is fixed and the intensity of the reflected light is mapped over the sample (see Figure \ref{fig:PrinciplesMaps} (c)) for a certain wavelength. For the IE mode the intensity of the reflected light is minimized by a 90\textdegree~alignment of the analyzer (one zone nulling condition) and $\Psi$ and $\Delta$ values are plotted. In IVASE mode $\Psi$ and $\Delta$ values are determined in dependence of the AOI and wavelength of the incident light. This information enables to determine the dielectric function of graphene.\\
An optical micrograph of graphene flakes with different layer numbers ranging from monolayer up to few-layer graphene and thin graphite is shown in Figure \ref{fig:PrinciplesMaps} (b). Employing the well established contrast method \cite{Abergel-Visibility-APL, Blake_MakingGrapheneVisible-APL}, the part of the left flake is identified as mono- and bilayer graphene. An IEI grey color plot of these graphitic sheets on \SiO/Si taken with a 20\texttimes~objective with AOI~=~42\textdegree~ a polarizer angle of 63.80\textdegree~ and an analyzer angle of 48.21\textdegree~ (compensator at 45\textdegree) using 552~nm light is displayed in Figure \ref{fig:PrinciplesMaps} (c). The grey scale reflects the real intensities captured by the CCD camera. Brighter regions therefore correspond to polarization changes of the reflected light that match the analyzer angle better than darker regions. Comparing Figure \ref{fig:PrinciplesMaps} (b) and (c) illustrates that the contrast in the IEI plot is appropriate to differentiate between graphene mono-, bi- and few layer. The sheets are better visible in the IEI image and both shape and layer number are clearly distinguishable by IEI. The height of the monolayer region was determined by AFM to be about 0.7~nm consistent with values for monolayers in literature \cite{Graf_nanoLett_SpatialResRaman}. Figure \ref{fig:PrinciplesMaps} (d) and (e) display IE $\Psi$ and $\Delta$ maps (AOI = 42\textdegree) of the framed region of Figure \ref{fig:PrinciplesMaps} (c).  The shape of the flake is unambiguously visible in the $\Psi$ as well as in the $\Delta$ map, however the $\Psi$ map gives a clearer signal for the monolayer region, whereas the $\Delta$ map displays a stronger one for the bilayer region.\par \noindent
\begin{figure}
\includegraphics{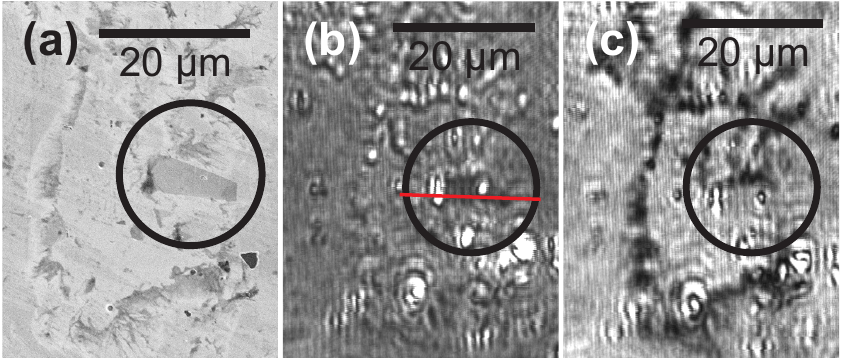}
\caption{(a) SEM image of few layer graphene on a GaAs substrate. The graphene  is centered in the circle and in the surroundings there are resist/tape residues. (b) and (c) are imaging ellipsometric intensity (IEI) plots of the same region. In (b) the contrast is optimized for the graphene layer such that the adhesive tape residues vanish, while in (c) the contrast of the immediate vicinity is enhanced.}
\label{fig:VisualizingOnGaAs}
\end{figure}
\begin{figure}
\includegraphics{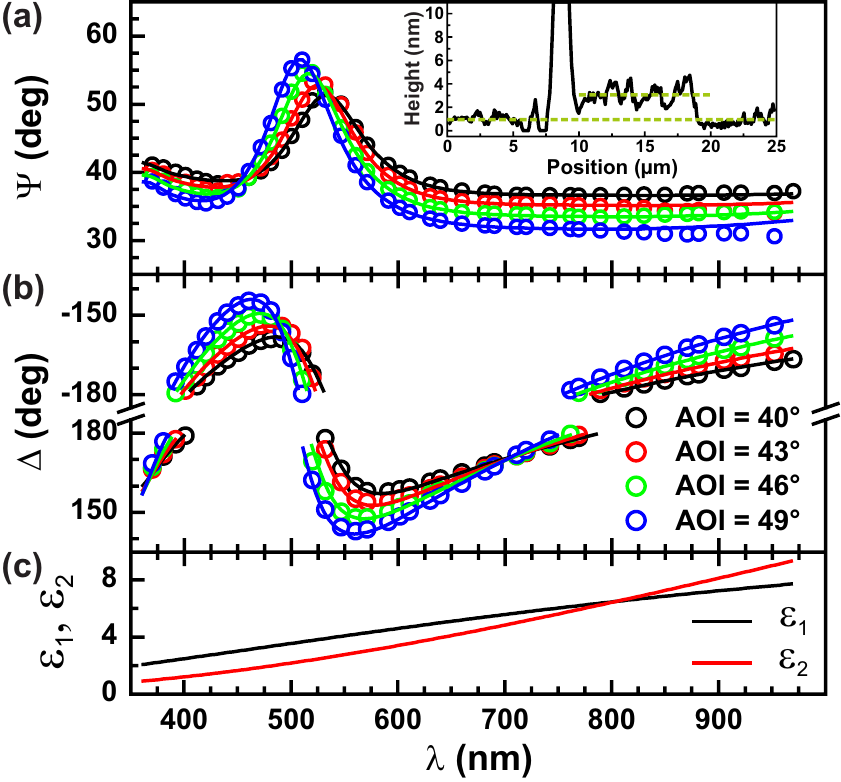}
\caption{(a) Wavelength dependent $\Psi$-angle of a graphene monolayer for different angles of incidence (AOI). Inset: Height profile of the flake on GaAs shown in Figure \ref{fig:VisualizingOnGaAs}. (b) Wavelength dependency of the $\Delta$-angle for the same AOI as in (a). (c) Dielectric coefficients $\epsilon_1$ and $\epsilon_2$ of graphene.}
\label{fig:Spectra}
\end{figure}

\noindent 
On substrates other than \SiO/Si graphene can hardly be detected with an optical microscope\cite{Wang_Nolte-ContrastClarity-APL}. As a demonstration of IEI's capability to determine shape and number of graphene layers on any flat substrate, we have investigated graphene on crystalline GaAs. In Figure \ref{fig:VisualizingOnGaAs} a graphene flake (marked with circles), deposited on the surface of GaAs, is imaged by (a) SEM and in (b) and (c) by IEI maps with different fixed angles between polarizer and analyzer at a wavelength of 532~nm. In the SEM image the contrast of the flake and the surrounding ''L-shaped'' resist/glue residues give strong contrast and only the straight edges of the flake enables to distinguish residues and graphene. The image shown in Figure \ref{fig:VisualizingOnGaAs} (b) was taken with an angle of 12.936\textdegree~ between polarizer and analyzer resulting in a large signal of the approximately 10~\textmu m long flake while the resist/glue residues are faint. Interestingly, a minor change of the angle to 19.095\textdegree~in Figure \ref{fig:VisualizingOnGaAs} (c) gives the opposite result: The contrast of the "L-shaped" residues is enhanced while the flake image fades away.\\
Under the assumption that the dielectric constants of graphene are constant and that the thickness of the layer is the only free fit parameter, the $\Psi$ map around the graphene sheet shown in Figure \ref{fig:VisualizingOnGaAs} (b) can be converted into a thickness map. The line profile along the red line is shown in the inset of Figure \ref{fig:Spectra} (a). The height of about~2.1~nm and the profile of this few-layer graphene flake is in good agreement with the corresponding AFM profile.\par
\noindent
Besides visualizing the flakes on various substrates, IVASE allows to determine optical properties of thin films. To extract the optical constants of graphene monolayers, several regions of interest (ROI) are carefully chosen from the IEI of the graphene monolayer displayed in Figure \ref{fig:PrinciplesMaps} (c). $\Psi$ and $\Delta$ values were measured for wavelengths between $\lambda = $~370~nm and $\lambda = $~952~nm at AOIs of 40\textdegree, 43\textdegree, 46\textdegree~ and 49\textdegree. The resulting values are averaged over all pixels of the CCD chip representing each ROI and collected in Figure \ref{fig:Spectra} (a) and (b). We made sure that the ROI was fully centered on a graphene flake as uncovered substrate contributes to the IVASE results. Consequently the investigated regions are centered on the graphene and the AOI was small enough ($\sim 45$\textdegree) to ensure that the measured spot is only placed on the graphene flake. As evident from Figure \ref{fig:Spectra} (a) and (b), both $\Psi$ and $\Delta$ depend on the used wavelength. The $\Psi$ values are less affected above $\lambda$ = 600~nm, while a maximum develops at about 500~nm. The $\Delta$ curves pass a local minimum below $\lambda$ = 575~nm and a maximum around $\lambda$ = 475~nm. The extremal values change slightly with the AOI. The experimental obtained values are in good agreement with the findings in reference \cite{GeimEllipsometry-PRB-2010}. The optical parameters have been modeled with the nanofilm\underbar~ep4model software based on the Fresnel coefficients for multilayered films and on the Drude model, provided by Accurion GmbH\cite{SpectroscopicEllipsometry-Tompkin,HandbookOfEllipsometry,Vaupel-JPhysChemC}. For the thickness of a graphene monolayer we have used 0.7~nm. The resulting dielectric coefficients $\epsilon_1$ and $\epsilon_2$ are plotted in Figure \ref{fig:Spectra} (c). Both coefficients increase with increasing wavelength $\lambda$.\\
Spectroscopy of $\Psi$ and $\Delta$ was also done on the few layer graphene flake on GaAs, shown in Figure \ref{fig:VisualizingOnGaAs}. The $\Delta$ values show a similar dependence on wavelength as found for graphene monolayers on \SiO, whereas the maximum of the $\Psi$ map seems to be shifted to wavelengths below 400~nm. The origin of these differences is unknown yet. They could either be caused by the different substrate materials or by increasing the layer number from monolayer (on \SiO) to few layer graphene (on GaAs). This will be explored in more detail in the future.\\
Modeling our data results in a similar dispersion as reported in reference \cite{Wang_Picometry} and show the same trend as reported in reference \cite{GeimEllipsometry-PRB-2010}. The origin of the quantitative difference in the extracted optical dispersion is not yet clear. One critical parameter in the simulations is the height of graphene. Here as well as in references \cite{ GeimEllipsometry-PRB-2010,Wang_Picometry} either literature values or AFM measured data have been used. A more independent way would be four-zone nulling spectroscopic ellipsometry, which is however out of the scope of this letter.\par\noindent
In conclusion, it has been shown that shape and number of layer of exfoliated graphene sheets can be determined on amorphous insulating \SiO~and crystalline semiconducting GaAs substrates by IE. From IVASE the optical properties can be extracted. This method enables to proof the prediction that the optical properties of graphene are dependent from the substrate. Further, the changes in the optical properties by including imperfections or lattice defects to the honeycomb lattice, e.g. by pattering into an antidot lattice \cite{JEroms-WL-NJP}, or the properties of different graphene edges could be explored with IE.\par\noindent
This work was supported by the DFG via GK 638 and GK 1570.


\end{document}